# Oscillating edge states in polariton topological insulators

Chunyan Li,[1,2,*] Yaroslav V. Kartashov[2]

[1]School of Physics, Xidian University, Xi'an, 710071, China
[2]Institute of Spectroscopy, Russian Academy of Sciences, 108840, Troitsk, Moscow, Russia
*Corresponding author: chunyanli@xidian.edu.cn

We show that slow time-periodic variation of the external magnetic field applied to polariton topological insulator based on honeycomb array of microcavity pillars with pronounced TE-TM splitting results in oscillations of the edge states along the boundary of the insulator that are accompanied by slow transverse drift of the center of mass of the edge state along the boundary. The oscillations and drift are due to time-periodic variation of the sign of the instantaneous group velocity of the edge state with selected central Bloch momentum triggered by the variation of the Zeeman splitting in the external magnetic field. These oscillations are also accompanied by periodic exchange of norm between two polarization components of the edge state. Oscillating edge states survive despite the fact that magnetic field periodically vanishes resulting in closure of the topological gap in the instantaneous spectrum of the system. The direction of the drift and amplitude of oscillations of the edge state strongly and nonmonotonically depend on its initial central Bloch momentum.

**Introduction**

Topological insulators are novel physical materials demonstrating properties sharply contrasting with properties of conventional insulators. While in their bulk topological insulators are characterized by the presence of the forbidden gap, they at the same time support at their edges localized states protected by the very topology of the system that cannot be destroyed without closing topological gap that makes them highly robust. Such unique properties of topological insulators attracted huge interest in diverse areas of physics and particularly in photonics (see recent reviews [1-3] and references therein). Unidirectional photonic edge states in systems with broken time-reversal symmetry were observed in gyromagnetic photonic crystals [4], arrays of coupled resonators [5], in Floquet topological insulators, where nontrivial topological phases emerge due to modulation of system parameters in time [6,7], and in many other systems. They were also predicted theoretically [8-12] and realized experimentally [13] in structured microcavities, where strong photon-exciton coupling leads to the formation of part-light part-matter polariton quasiparticles. Due to their excitonic component, such quasiparticles are sensitive to the external magnetic field. In this system nontrivial topological phase in specially patterned microcavity emerges in the presence of TE-TM splitting of cavity modes (emulating spin-orbit coupling) and constant external magnetic field, whose sign determines circulation direction of the unidirectional topological edge states. Rich inherent nonlinear interactions between polaritons with different polarizations allow to use resonant pumping to induce antichiral topological currents in these systems [14]. Such currents can also be created by applying sublattice-dependent Zeeman splitting to polariton condensates with spin-orbit coupling [15]. In general, the combination of the effects of TE-TM splitting and external magnetic fields allow to implement interesting topological spin-filtering schemes with polaritons [16]. More details on implementation of topological insulators in polariton microcavity systems can be found in recent review [17].

Additional opportunities for control of evolution dynamics of excitations in topological systems arise when underlying topological potential landscapes are additionally modulated in evolution variable (such as time in polaritonic systems or propagation distance in photonic ones). Such modulations bring a host of nontrivial dynamical phenomena even in trivial lattices, among which one can mention dynamic localization, inhibition of tunneling, rectification, and controllable reshaping of the wavepackets [18]. In polaritonic systems temporal modulations of the potential energy landscape in microcavity can be used to control group velocity of the edge states [19], it induces resonant edge state switching [20] or Rabi-like oscillations [21]. In addition, inclusion of the transverse potential gradient along the edge of polariton topological insulator induces anomalous Bloch oscillations upon which the wavepacket can move in the direction transverse to the gradient [22]. Even temporal modulations of nonresonant pump in polariton systems may allow to excite different self-sustained stationary or dynamically evolving structures in the same microcavity [23].

Nontrivial topological phases can also be induced by spatially inhomogeneous magnetic fields, as demonstrated in atomic systems. Such fields may create a new type of Zeeman lattice, in which potential maxima for one spin component coincide with potential minima in other component [24]. Under the action of spin-orbit coupling truncated Zeeman lattices support topological edge states [25] and may allow the formation of topological Bragg solitons [26]. Nevertheless, the behavior of unidirectional topological edge states under time-periodic variations of the magnetic field was not studied neither in atomic, nor in polaritonic systems.

In this work, we study unusual regime of evolution of the edge states in polariton topological insulators. We show that periodic variation of the external magnetic field induces oscillations of the edge states, that are accompanied by unexpected linear drift of the center of mass of the state along the edge, whose rate is determined by the initial Bloch momentum. In the course of such oscillations, the edge states remain bound to one edge and survive despite the fact that magnetic field periodically vanishes. Notice that in contrast to topological Floquet systems with sufficiently fast temporal variations of parameters, where edge states emerge namely due to such temporal variations [27], here we consider very slow, adiabatic variation of the Zeeman splitting under the action of time-dependent magnetic field, so formally edge states in our case follow adiabatic variations of parameters of the system.

**The Model**

We describe the evolution of polaritons in structured microcavity using normalized coupled Schrödinger equations for spin-positive ($\psi_+$) and spin-negative ($\psi_-$) components of the polariton wavefunction written in circular polarization basis [11]:

$$i\frac{\partial \psi_\pm}{\partial t} = -\frac{1}{2}\Delta \psi_\pm + \beta(\partial_x \mp i\partial_y)^2 \psi_\mp \pm \Omega(t)\psi_\pm + \mathcal{R}(x,y)\psi_\pm, \quad (1)$$

where $\Delta = \partial_x^2 + \partial_y^2$, $x, y$ are scaled to $x_0 = 1\ \mu$m transverse coordinates, $t$ is the scaled evolution time, the parameter $\beta$ characterizes spin-orbit coupling strength, Zeeman splitting in time-periodic external magnetic field is described by $\Omega(t) = \Omega_0 \cos(\omega t)$ with amplitude $\Omega_0$ and frequency $\omega$. The potential energy landscape describing honeycomb ribbon with zigzag-zigzag edges in structured microcavity is set by $\mathcal{R}(x,y) = -p\sum_{m,n} e^{-[(x-x_m)^2 + (y-y_n)^2]/d^2}$, where $p$ is the potential depth and $d = 0.4$ is the width of pillars located in the nodes $(x_m, y_n)$ of the honeycomb grid with spacing $a = 1.4$ between nodes [see Fig. 1(a) for $-\mathcal{R}$ profile]. All energy parameters (such as the potential depth and the Zeeman splitting amplitude) are scaled to $\varepsilon_0 = \hbar^2/mx_0^2 \approx 0.35$ meV, where $m \sim 10^{-34}$ kg is the effective polariton mass, evolution time is scaled to $t_0 = \hbar \varepsilon_0^{-1} \approx 1.9$ ps. Potential depth $p = 8$ corresponds to $\sim 2.8$ meV. The frequency $\omega \sim 10^{-3} - 10^{-2}$ is selected to guarantee adiabatic evolution. Further we set $\beta = 0.3$ and $\Omega_0 = 0.8$.

We do not take into account in model (1) the intrinsic losses in polariton microcavity because the very existence of the edge states in such systems is not connected with losses, and because such losses can be compensated by suitable external pump [28,29]. Although in experiments such a pump may act differently on edge and bulk states, leading to potential differences in absorption/gain for different parts of the wavepacket that may slightly impact its evolution dynamics, in modern samples with sufficiently long polariton lifetimes [17] one should be able to observe at least one period of oscillations described below.

**Result and discussion**

In the case, when external magnetic field is constant and does not change with time, the Bloch eigenmodes of the Eq. (1) can be obtained as $\psi_\pm(x,y) = \varphi_\pm(x,y)e^{iky - i\varepsilon(k)t}$, where $\varepsilon(k)$ is the energy, $k$ is the Bloch momentum along the $y$-axis in the first Brillouin zone of width $K = 2\pi/Y$, where $Y = 3^{1/2}a$ is the $y$-period of structure, and $\varphi_\pm(x,y) = \varphi_\pm(x, y+Y)$, i.e. zero (periodic) boundary conditions are applied here in the $x$ ($y$) direction in the calculation. The dependence of energy $\varepsilon$ on momentum $k$ is presented in Fig. 1(a) and 1(b) for two opposite values of the Zeeman splitting $\Omega = +\Omega_0$ and $\Omega = -\Omega_0$. Only two lowest bands in the spectrum are shown. The green (red) branches denote the edge states residing at the left (right) edges of the array. One can see that the inversion of the sign of the magnetic field does not change the spectrum qualitatively except for the fact that the states residing on a given edge of the structure correspond to branches with opposite slopes for opposite $\Omega$ values. The profiles of the edge states from the left edge corresponding to the same value of the Bloch momentum $k$ are also presented in Fig. 1(a) and 1(b). One can see that edge states from the same edge have different dominating components: while at positive $\Omega$ the $\psi_-$ component clearly dominates (has larger amplitude and norm) over $\psi_+$ one, the situation is reversed for negative $\Omega$, when $\psi_+$ component becomes dominating. Since in the process of periodic adiabatic variation of $\Omega$ the state remains predominantly near the same edge (see discussion of dynamics below), one may assume that the variation of $\Omega$ will be accompanied by the exchange of norm between $\psi_+$ and $\psi_-$ components in the edge state.

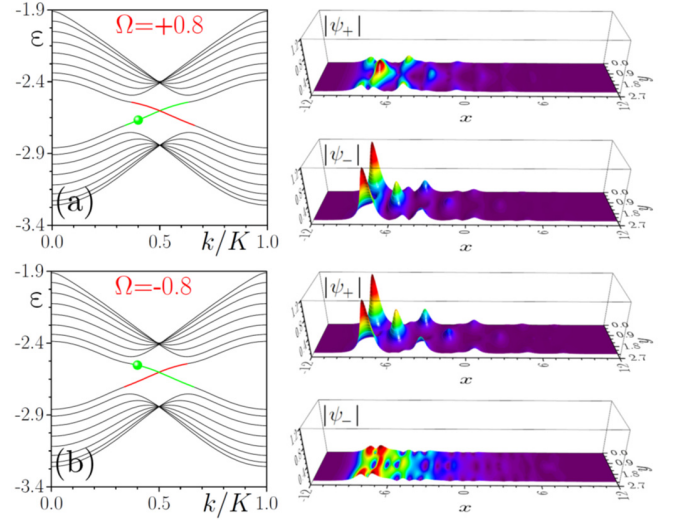

Fig. 1. Energy-momentum diagram under constant Zeeman splitting $\Omega = +\Omega_0$ (a) and $\Omega = -\Omega_0$ (b). Green (red) lines correspond to unidirectional topological edge states localized at the left (right) boundary of the potential. Examples of modulus distributions of component $\psi_+$ ($\psi_-$) denoted by green spheres in (a) and (b) at $k = 0.4K$ are displayed in top (bottom) row of each right panel. In all cases $\beta = 0.3$.

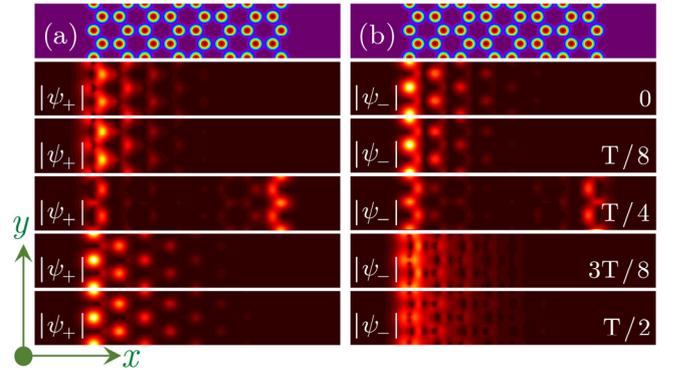

Fig. 2. Top row shows two periods of $-\mathcal{R}$ landscape with zigzag-zigzag edges. Second to sixths rows show examples of modulus distributions in $|\psi_+|$ (a) and $|\psi_-|$ (b) components of edge states with $k = 0.4K$ formally corresponding to green stars in Fig. 3(a-c) and (e, f).

Due to slow adiabatic variation of the external magnetic field $\Omega(t)$ one can assume that instantaneous spectra of the system calculated in different moments of time $t$ may explain main features of the dynamical evolution of the edge states. The instantaneous eigenmodes of the ribbon with zigzag-zigzag edges depicted in Fig. 2(a) were calculated for different values of $\Omega$ corresponding to different characteristic moments of time $t$. Instantaneous spectra calculated at dif-ferent moments of time $t$ within one period $\mathrm{T}=2\pi/\omega$ of oscillation of the magnetic field are presented in Fig. 3. Only two lowest bands are shown. As before, green (red) branches represent topological edge states localized at the left (right) edge of the ribbon, blue branches correspond to trivial edge states, black curves correspond to bulk modes. When magnetic field is sufficiently strong the topological gap is wide that guarantees confinement of the mode near the given edge. When magnetic field $\Omega(t)$ vanishes at $t=\mathrm{T}/4$ and $3\mathrm{T}/4$ the gap closes and topological edge states [green and red stars in Fig. 3(a)] transform into nontopological ones [blue stars in Fig. 3(c) and 3(i)], that is accompanied by corresponding transformation of modal shapes, examples of which at $k=0.4K$ are shown in Fig. 2. We should mention here that in sharp contrast to the topological one, the non-topological edge states reside simultaneously at two boundaries of the array (see instantaneous mode at $t=\mathrm{T}/4$ in Fig. 2 for example). Even though eigenvalue solver returns nontopological edge states with both edges being populated (because at $\Omega=0$ time-reversal symmetry of the system is not broken), this does not mean that upon dynamical evolution polaritons can couple to the opposite edge – instead, they remain in the vicinity of excited edge at all moments of time (see Fig. 4 with dynamics below). Variation of the magnetic field leads to time-periodic variation of magnitude and sign of the group velocity $\partial\varepsilon/\partial k$ of the edge states at a given edge (red and green branches reverse their order in the moments $t=\mathrm{T}/4$ and $3\mathrm{T}/4$), thus one can expect that propagation direction of such states will change with time. Spectra at two specific moments of time, when edge states at $k=0.4K$ and $k=0.6K$ change their propagation direction, since at these moments and for these momentum values $\partial\varepsilon/\partial k$ becomes zero, are shown in Fig. 3(d) and 3(h). It should be also mentioned that while spectra in time moments symmetric with respect to $t=\mathrm{T}/4$ or $3\mathrm{T}/4$ look qualitatively similar (compare spectra at $t=\mathrm{T}/8$ and $t=3\mathrm{T}/8$) the relative strengths of $\psi_+$ and $\psi_-$ spin components in edge state residing at a given boundary notably change (thus, $\psi_-$ component dominates at $t<\mathrm{T}/4$, while $\psi_+$ component dominates at $t>\mathrm{T}/4$), indicating on the fact that variation of the magnetic field also causes periodic exchange of norm between spin components, as was suggested upon discussion of Fig. 1.

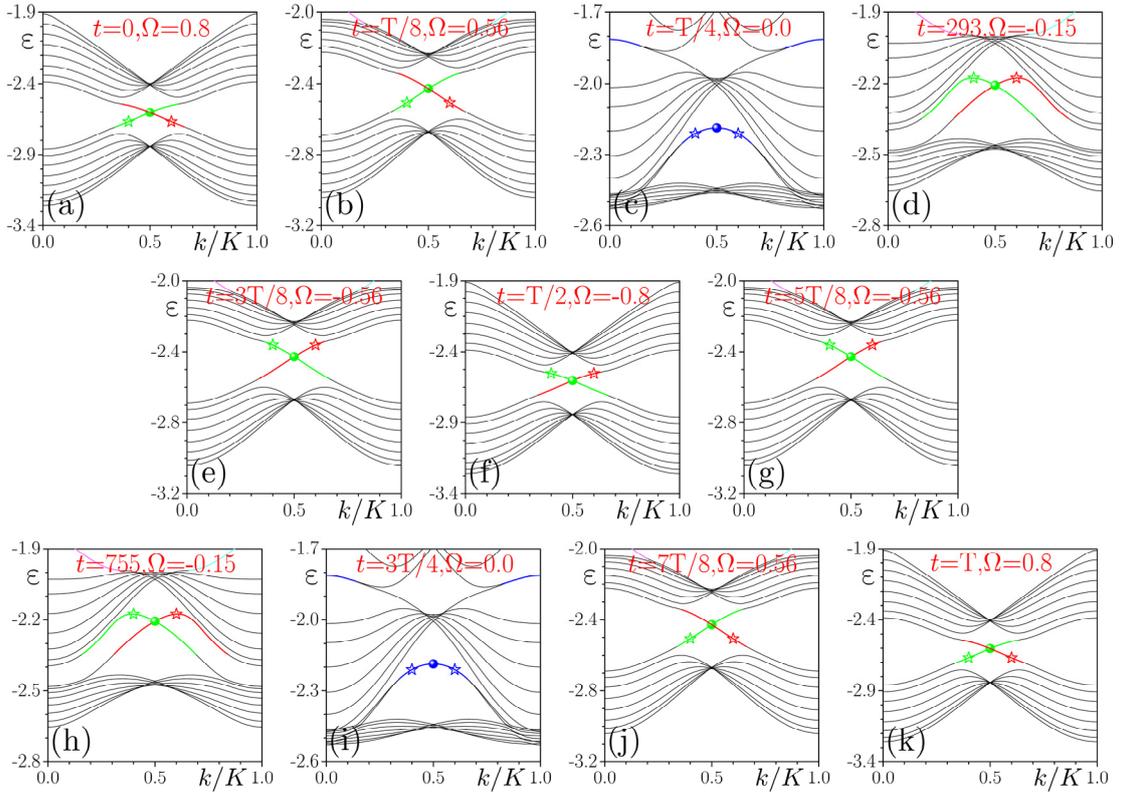

Fig. 3. Examples of instantaneous spectra $\varepsilon(k)$ of the system in different moments of time. Red and green lines corresponding to unidirectional topological edge states, blue lines highlight degenerate non-topological edge states. Colored stars and spheres indicate energies of dynamically evolving edge states obtained using Eq. (2), (3). $|\psi_\pm|$ distributions in stationary edge states formally corresponding to green stars in (a-c) and (e-f) are displayed in Fig. 1. Here $\omega=0.006$.

To study the dynamics of the edge states under time-periodic magnetic field we now launch into system the states calculated at $t=0$ with wide envelope $\psi_\pm(x,y)|_{t=0}=A(y)\varphi_\pm(x,y)e^{iky}$, where $A(y)=e^{-y^2/w_d^2}$ and $w_d=30$. Figure 4 shows evolution dynamics for three different Bloch momenta. The dynamics depicted in Fig. 4 reveals steady oscillations of the edge states in accordance with dynamically varying group velocity $\partial\varepsilon/\partial k$ in time-periodic magnetic field, that is however accompanied by unexpected linear drift of the center of mass of the edge state, when it is calculated in time moments $t=n\mathrm{T}/2$ (see dashed yellow lines indicating direction of the drift). Notice that the rate of this drift is substantially smaller than the group velocity that this edge state would have in the presence of constant magnetic field $\Omega\equiv\Omega_0$. In all cases, the edge state remains in the vicinity of the left edge, where it was launched and does not transform into nontopological edge states, or diffract into the bulk, even in the time moments when magnetic field vanishes. For example, the state at $k=0.4K$ initially moves in the positive direction of the $y$-axis, since its group velocity $\partial\varepsilon/\partial k>0$. Beyond $t=\mathrm{T}/4$, when group velocity becomes negative the state starts moving in the negative direction of the $y$-axis. Next reversal of propagation direction occurs around $t=3\,\mathrm{T}/4$, when magnetic field again changes its sign. Remarkably, the direction of the average drift of the edge state depends on initial Bloch momentum: it is opposite for $k=0.4K$ and $k=0.6K$, while for $k=0.5K$ the drift is absent. The edge state also slowly diffracts along the $y$-axis (compare $|\psi_-|$ distributions at $t=0$ and $t=2\,\mathrm{T}$), with diffraction rate determined by the average dispersion $\partial^2\varepsilon/\partial k^2$. Notice that while simulations presented here are based on sufficiently narrow ribbon with 4 honeycomb cells in the $x$ direction (see Fig. 2), we checked that persistent oscillations can also be observed in much wider structures containing up to 24 cells in the $x$ direction.

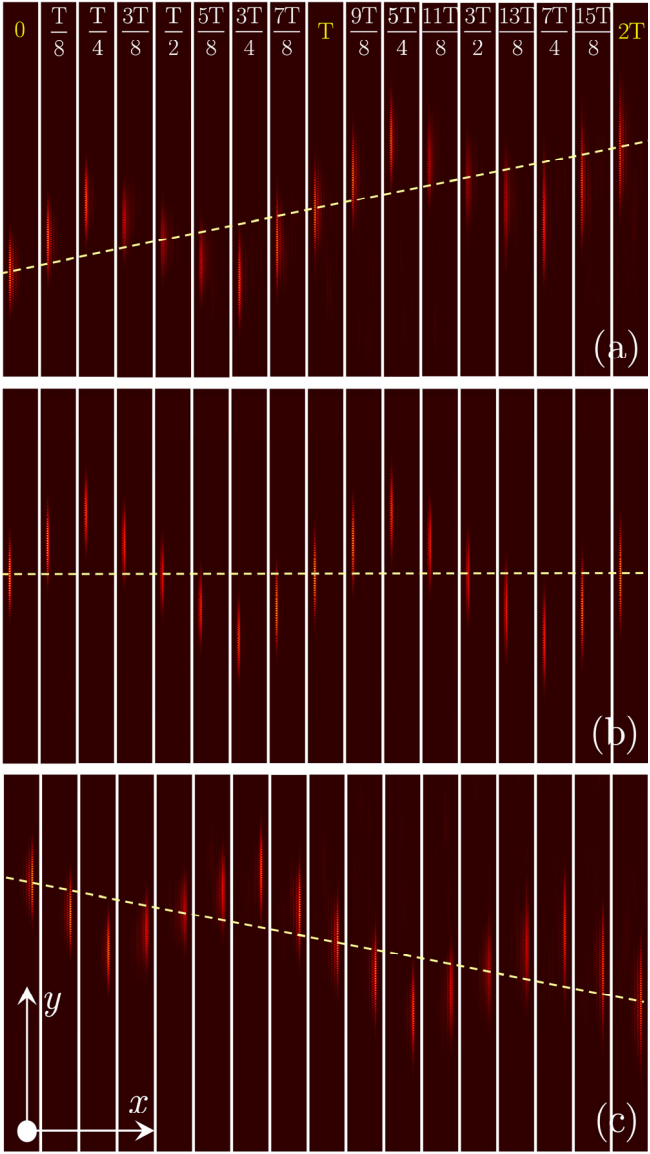

Fig. 4. $|\psi_-|$ distributions in different moments of time indicated on the plots. Input states in (a-c) are edge states corresponding to green star ($k=0.4K$), green sphere ($k=0.5K$), and red star ($k=0.6K$) in Fig. 3(a) with a wide Gaussian envelope. Yellow dashed line highlights linear drift calculated using edge state positions at $t=n\mathrm{T}/2$.

To characterize oscillations in Fig. 4 we calculate center of mass position $y_c$ for the edge state and its instantaneous energy $\varepsilon_t$:

$$y_c = N^{-1}\iint y\mathbf{\Psi}_t^\dagger\mathbf{\Psi}_t dxdy,$$
$$\varepsilon_t = (i/dt)\ln\left[\iint\mathbf{\Psi}_t^\dagger\mathbf{\Psi}_{t+dt}dxdy\right], \quad (2)$$

where we introduced spinor wavefunction $\mathbf{\Psi}_t=(\psi_+,\psi_-)^\mathrm{T}$ at time moment $t$, $N=\iint\mathbf{\Psi}_t^\dagger\mathbf{\Psi}_t dxdy$ is the total norm normalized to 1 for $\mathbf{\Psi}_{t=0}$, and $dt$ is the time step in modeling of evolution. Calculation of projections

$$\mathrm{P}_{k'} = \iint\mathbf{\Psi}_t^\dagger\mathbf{\Psi}_{t=0,k'}dxdy \quad (3)$$

of the evolving polariton wavefunction $\mathbf{\Psi}_t$ corresponding to input edge state with momentum $k$ on all eigenmodes $\mathbf{\Psi}_{t=0,k'}$ of the system with various momenta $k'$ clearly shows that central momentum of the state is conserved upon evolution in time-periodic magnetic field. Taking into account this fact, calculated instantaneous energy $\varepsilon_t$ allows to trace the location of the evolving wavepacket $\mathbf{\Psi}_t$ in the instantaneous energy spectrum $\varepsilon(k)$ depicted in Fig. 3 (see, for example, green and red stars showing spectral positions of the evolving wavepackets from Fig. 4(a) and 4(c)]. The dependencies $y_c(t)$ and $\varepsilon_t(t)$ are presented in Fig. 5. Their inspection in conjunction with spectrum in Fig. 3 reveals that wavepacket at $k=0.4K$ acquires zero group velocity $\partial\varepsilon/\partial k$ slightly after $t=\mathrm{T}/4$ and before $t=3\,\mathrm{T}/4$ [see left blue stars in Fig. 3(c) and 2(i)], therefore $\partial\varepsilon/\partial k$ remains positive over larger time interval on one period $\mathrm{T}$. The opposite happens for the wavepacket with $k=0.6K$ that has negative velocity during larger time interval. Namely this is the reason for opposite drift directions of these edge states because they move with positive and negative group velocities on slightly different time intervals within one period. In contrast, for state with $k=0.5K$ group velocity becomes zero exactly in the time moments $t=\mathrm{T}/4$ or $3\,\mathrm{T}/4$, so that its velocity acquires positive and negative values on exactly equal time intervals, and no net drift is observed. Figure 5(b) reveals exactly periodic evolution of energy of the edge states, it also shows that there is periodic exchange of norm between two polarization components. Figure 5(a) allows to calculate the amplitude of oscillations $A_y=\max(y_c)-\min(y_c)$ of the edge state and its shift $S_y$ over one period $\mathrm{T}$. They both monotonically decrease with increase of modulation frequency $\omega$ [Fig. 6(a) and 6(b)]. At the same time $S_y$ is a nonmonotonic function of momentum $k$ [Fig. 6(c)] reaching maximal values sufficiently far from $k=0.5K$. Instead, the amplitude of oscillations is maximal at $k=0.5K$ and it decreases away from this point [Fig. 6(d)].

As one can see from Fig. 5, for small frequencies $\omega\sim0.006$ of variation of the magnetic field, the amplitude of wavepacket oscillations practically does not change with time (i.e. it is the same for subsequent oscillation periods). In this regime, no radiation is visible from the wavepacket – its oscillations are persistent and it only slowly broadens in space due to dispersion. Nevertheless, radiation becomes visible for frequencies $\omega>0.02$ where it starts affecting $y_c(t)$ dependencies.

**Conclusions**

In conclusion, we have reported on unusual oscillations of the topological edge states arising from time-periodic magnetic field that are accompanied by linear drift of the edge state along the edge that can be controlled by its momentum, which significantly affects the direction and the magnitude of the drift. These oscillations survive even though magnetic field periodically vanishes. The scheme suggested here offers intriguing opportunities for control of excitation dynamics in topological systems with slowly varying in evolution variable parameters and it can be potentially extended also to spinor atomic systems.

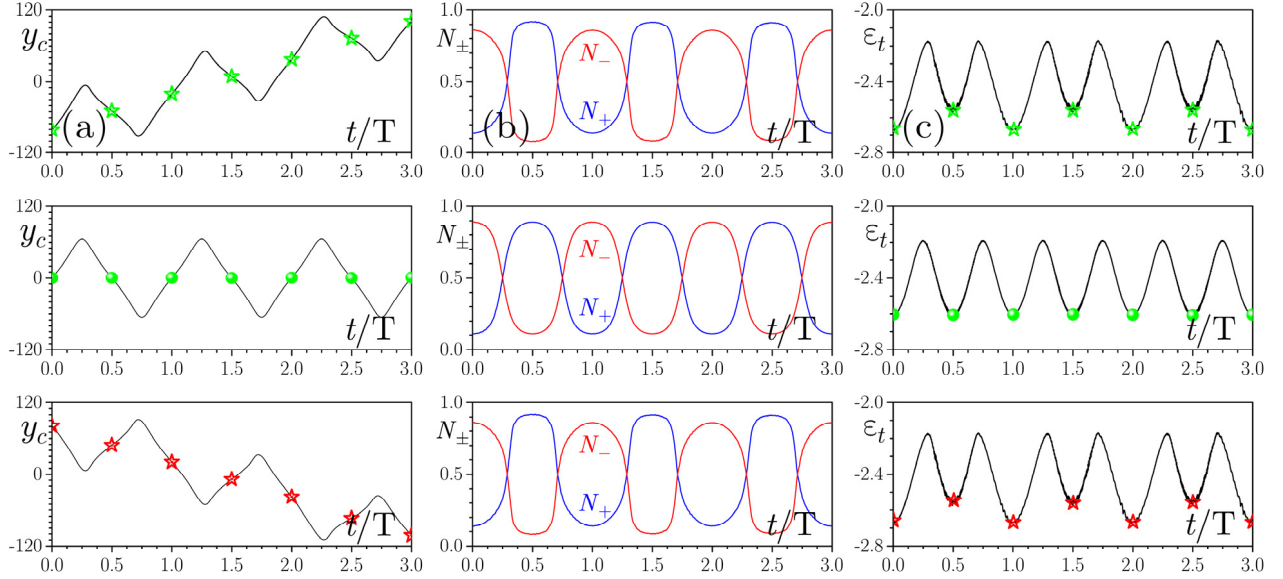

Fig. 5. Center of mass of the wavepacket $y_c$ (a), norms of two polarization components (b), and instantaneous energy (c) versus time for $k=0.4K$, $0.5K$, and $0.6K$. Colored symbols in (a) and (b) correspond to the time moments $t=nT/2$, $n=0,\ldots,6$.

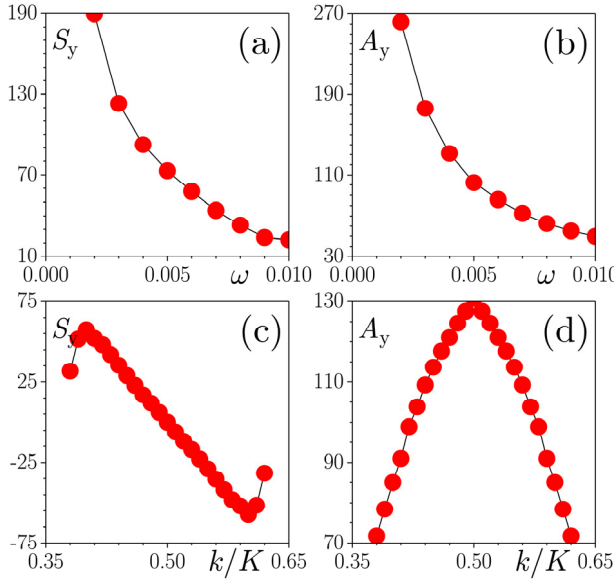

Fig. 6. (a),(b) Shift and amplitude of $y$-oscillations of the edge states as a function of $\omega$ for $k=0.4K$. (c), (d) Shift and amplitude of oscillations as a function of momentum $k$ for fixed frequency $\omega=0.006$.

### Acknowledgements

The support of this study by National Natural Science Foundation of China (NSFC) (11805145); China Scholarship Council (CSC) (202006965016); and Research project FFUU-2021-0003 of the Institute of Spectroscopy of the Russian Academy of Sciences is acknowledged.